\documentclass{iopart}

\usepackage{graphicx}
\usepackage{graphics}
\expandafter\let\csname equation*\endcsname\relax
\expandafter\let\csname endequation*\endcsname\relax
\usepackage{amsmath}
\usepackage{bm}
\usepackage{hyperref}
\usepackage{iopams}
\usepackage{amsthm}
\usepackage{float}

\usepackage[dvipsnames]{xcolor}

\begin{document}

\title[Resonant jumps induced by stationary tidal perturbation: a two-for-one deal]{Resonant jumps induced by stationary tidal perturbation: a two-for-one deal}

\author{Priti Gupta} 
\address{Department of Physics, Kyoto University, Kyoto 606-8502, Japan}
\ead{priti.gupta@tap.scphys.kyoto-u.ac.jp}

\author{Takafumi Kakehi} 
\address{Yukawa Institute for Theoretical Physics,Kyoto University, Kyoto 606-8502, Japan}
\address{Center for Gravitational Physics and Quantum Information, Yukawa Institute for Theoretical Physics,Kyoto University, Kyoto 606-8502, Japan}
\ead{takafumi.kakehi@yukawa.kyoto-u.ac.jp}

\author{Takahiro Tanaka} 
\address{Department of Physics, Kyoto University, Kyoto 606-8502, Japan}
\address{Center for Gravitational Physics and Quantum Information, Yukawa Institute for Theoretical Physics,Kyoto University, Kyoto 606-8502, Japan}
\ead{t.tanaka@tap.scphys.kyoto-u.ac.jp}

\begin{abstract}
Extreme-mass-ratio inspirals (EMRIs) are promising target sources for space-based interferometers such as LISA, Taiji, and Tianqin. Depending on the astrophysical environment, such as close perturbers or an accretion disk, EMRI orbital evolution may deviate from the predictions of general relativity in  vacuum. In particular, we focus on  the resonance jumps, {\it i.e.}, the changes of the conserved quantities induced by a stationary perturbation to the background Kerr geometry. Using Hamiltonian formulation, we provide a closed relation between the jump in Carter constant and that in the axial component of angular momentum. It is also shown that the obtained relation is consistent with the fitting formulae computed for the tidal resonance in previous works.  
\end{abstract}

\section{Introduction}
\label{sec:intro}
LIGO interferometers detected gravitational waves (GWs) for the first time in 2015, leading to the first direct observation of GWs~\cite{LIGOScientific:2016aoc}. With the space-based interferometers (such as LISA~\cite{amaroseoane2017laser}, Taiji~\cite{Ruan:2018tsw}, and Tianqin~\cite{TianQin:2020hid}), we can detect sources emitting GWs at mHz frequencies. There are a number of promising sources for LISA, including extreme-mass-ratio inspirals (EMRIs), which involve stellar-mass objects spiraling into supermassive black holes (SMBHs)~\cite{Babak:2017tow,berry2019unique,Barausse:2020rsu}. Given the extreme mass ratio, EMRIs will spend months to years in the observable band before the plunge, allowing us to map strong gravity spacetime and test the black hole geometry with great precision. 
High precision theoretical prediction of the gravitational waveform is an urgent task, and a lot of efforts have been made~\cite{Barack:2018yvs,Pound:2021qin}. 

Generic relativistic bound orbits around massive BHs have three frequencies, {\it i.e.}, the radial $\omega_{r}$, polar $\omega_\theta$, and azimuthal $\omega_\phi$ frequencies. Due to the self-force~\cite{Mino_1997,Quinn_1997}, these frequencies evolve adiabatically as binary shrinks. 
The evolution track of an EMRI passes through several self-force resonances, unless it takes a quasi-circular or equatorial orbit. During such resonances, radial and polar frequencies become commensurate, {\it i.e.}, $n\,\omega_{r}+k\,\omega_\theta=0,$ where $n,k$ are integers~\cite{PhysRevLett.109.071102}. EMRIs can be significantly affected by a perturbation at resonance points. At the resonance, a ``jump" (depending on the orbital phase) is induced in the ``constants of motion'' (orbital energy $E$, axial angular momentum $L_z$, and Carter constant $Q$), which alters the post-resonance orbital evolution~\cite{Flanagan:2012kg,PhysRevD.94.124042}.

In a similar manner, when we have an external perturber, such as a neighboring third body, a tidal gravitational  perturbation is added to the background spacetime of the EMRI, giving rise to \textit{tidal resonances}\footnote{Generically, when the EMRI's three orbital frequencies are commensurate with those of the perturber, a tidal resonance occurs. This paper treats the perturber as static, so its orbital frequencies are irrelevant to resonance conditions.}~\cite{byh} when $n\,\omega_{r}+k\,\omega_\theta+m\,\omega_\phi=0$. It has been shown that EMRIs may encounter multiple tidal resonances and may dephase the waveform significantly depending on the magnitude of the tidal perturbation as well as on the orbital parameters and phase. A slowly changing tidal perturbation will result in jumps in the azimuthal angular momentum and Carter constant, and fitting formulae for different resonance combinations have already been developed~\cite{PaperI,Gupta:2022fbe}. It may, however, be necessary to evaluate a broader range of resonances in the future. Recently, a simpler method of computing the resonant jump for angular momentum was proposed~\cite{Silva2022}. However, this method does not directly apply to the jump in the Carter constant. In this paper, we show that there is a simple relation between the two jumps caused by a stationary external perturbation, which is not restricted to the tidal perturbation with $|m|\leq 2$. This relation allows us to deduce the jump in the Carter constant from the jump in the angular momentum. 


\section{Brief review of the Hamiltonian approach}
Self-force motion can be understood using a perturbed Hamiltonian expressed in terms of action-angle variables~\cite{PhysRevD.78.064028}. Here, we argue that the same method as for the conservative part of the self-force, whose existence was first pointed out in~\cite{Isoyama:2013yor} and 
very recently confirmed by~\cite{Nasipak:2022xjh}, can also be used in the case of external perturbation. In particular, we consider a tidal perturbation added to the background Kerr spacetime. The full metric is given by 
\begin{align}
g_{\mu\nu}=g_{\mu\nu}^{(0)}+h_{\mu\nu}\,,
\end{align}
where 
\begin{align}
g^{(0)}_{\mu\nu}dx^\mu dx^\nu=& -\left(1-\frac{2M r}{\Sigma}\right)dt^2-\frac{4Mar \sin^2\!\theta}{\Sigma}dt\, d\phi +\frac{\Sigma}{\Delta}dr^2\cr
 & +\Sigma\, d\theta^2+\left(r^2+a^2+\frac{2Ma^2 r}{\Sigma}\sin^2\!\theta \right)\sin^2\! \theta\, d\phi^2\,,
\end{align}
with 
\begin{align}
\Sigma\equiv r^2+a^2 \cos^2\!\theta,\qquad \Delta\equiv r^2-2Mr+a^2\,,
\end{align}
is the infinitesimal line element of the Kerr spacetime, and
$h_{\mu\nu}$ represents a stationary external perturbation, which is given independently of the particle motion. 
The perturbed metric admits a stationary Killing vector 
$$
(\partial_t)^\mu-\Omega (\partial_\phi)^\mu\,,$$
with $\Omega$ being the frequency of the perturber. We consider the geodesic motion of a particle in this perturbed spacetime, which is governed by the following Hamiltonian 
\begin{align}
 H=H^{(0)}+H^{(1)}
  =\frac{1}{2} g_{\mu\nu}\left(x(\tau)\right) u^\mu(\tau) u^\nu(\tau)\,, 
\end{align}
where $H^{(0)}$ and $H^{(1)}$ describe the contributions coming from 
$g_{\mu\nu}^{(0)}$ and $h_{\mu\nu}$, respectively. 
Our focus is on the change rates of $E, L_z$ and $Q$, which are defined by
\begin{align}
 &E \equiv-\mu u_t\,, \\
 &L_z \equiv\mu u_z\,, \\
 &Q \equiv \mu^2 (u_\theta^2 +a^2\cos^2\! \theta (1-u_t^2)+\cot^2\!\theta u_\phi^2)\,,
\end{align}
with $u^\mu$ being the four-velocity of the particle~\cite{PhysRevD.78.064028}. 
Additionally, the mass of the particle $\mu^2$ is also treated as a constant of motion such that
\begin{align}
  \mu^2=-g^{\mu\nu}p_\mu p_\nu\,.
\end{align}
where $p_\mu$ is the conjugate momenta of the particle. The complete set of quantities $\{E, L_z, Q, \mu^2\}$ is denoted by $\{I_a\}$. For the following formal discussion, we use a set of action-angle variables, $(w^a, J_a)$, instead of the pair of the Boyer-Lindquist coordinates and the specific four momentum $(x^\mu, p_\mu)$. The relation between these two sets of variables is given by 
\begin{align}
p_\mu=\frac{\partial W}{\partial x^\mu} (x^\nu,J_a)\,,\\
w^a=\frac{\partial W}{\partial J_a}(x^\nu,J_a)\,,
\end{align}
using the generating function $W(x^\nu,J_a)$ defined as
\begin{align}
W(x^\mu,J_a)=-J_t t+J_\phi \phi \pm \int^r dr' \frac{\sqrt{V_{r}(r^\prime)}}{\Delta} \pm \int ^\theta d\theta' \sqrt{V_{\theta}(\theta^\prime)}\,,
\end{align}
where the functions $V_r$ and $V_\theta$ are given by 
\begin{align}
 & V_r (r,E,L_z,Q,\mu^2)=\left(E(r^2+a^2)-a L_z\right)^2
   -\Delta \left[r^2\mu^2+(L_z-a E)^2+Q\right]\,,\cr
 & V_\theta(\theta,E,L_z,Q,\mu^2)=Q-\cos^2\!\theta\left[a^2(\mu^2-E^2)+\frac{L_z^2}{\sin^2\!\theta}\right]\,.
\end{align}
The expression for $W(x^\mu,J_a)$ contains $E,L_z,Q$ and $\mu^2$, which are to be understood as functions of $J_a$ determined by solving 
\begin{align}
 & J_t=-E\,,\quad J_\phi=L_z\,,\quad\cr
 & J_r=\frac1{2\pi}\oint\frac{\sqrt{V_r(r,E,L_z,Q,\mu^2)}}{\Delta(r)} dr\,,\cr
 & J_\theta=\frac1{2\pi}\oint\sqrt{V_\theta(\theta,E,L_z,Q,\mu^2)} d\theta\,. 
\end{align}
Here, $\oint$ means the integral over one period of oscillation of the integration variable, and the signature of the square root is chosen to be positive (negative) when the integration variable increases (decreases). 

The change rate of $J_a$ is given by the corresponding 
Hamiltonian equation of motion, 
\begin{align}
 \frac{dJ_a}{d\tau}=-\frac{\partial H}{\partial w^a}
 =-\frac{\partial H^{(1)}}{\partial w^a_{\rm i}}\,, 
 \label{eq:Jdot}
\end{align}
where $w^a_{\rm i}$ is the initial value of $w^a=\{w^t, w^r,w^\theta,w^\phi\}$. 
In the second equality we used the facts that $H^{(0)}$ is written solely in terms of $J_a$, and $w^a$ for the background geodesic motion is expressed as $w^a=\omega^a \tau +w^a_{\rm i}$ with $\omega^a$ being the angular velocity of the variable $w^a$. 
For the non-resonant orbit, the average of the right-hand side of Eq.~\eqref{eq:Jdot} vanishes because the long time average of $H^{(1)}$ (denoted by $\left\langle H^{(1)}\right\rangle$), should be independent of the choice of the initial phase. Hence, $\left\langle H^{(1)}\right\rangle$ is also a function of $J_a$ only, as in the case of $H^{(0)}$. By contrast, when the resonant condition is satisfied, {\it i.e.}, when 
\begin{align}
  n \omega^r+k \omega^\theta + m (\omega^\phi+\Omega \omega^t)=0\,,
  \label{eq:resonanceCondition}
\end{align}
holds for some set of small non-trivial set of integers $\{n, k, m\}$, 
the averaging must be treated differently. We refer to such a case as the resonance combination {$n:k:m$}, although in general neither the ratio of frequencies nor that of their inverses takes a simple integer ratio, it is the ratio of the coefficients in Eq.~\eqref{eq:resonanceCondition} that takes the simple integer ratio $n:k:m$.   
In such a case, the background geodesics taking different initial values of 
\begin{align}
  \Delta w\equiv n w^r+k w^\theta + m (w^\phi+\Omega w^t)\,,
\end{align}
modulo $2\pi$, are distinguishable. Hence, the average value $\left\langle H^{(1)}\right\rangle$
depends on $\Delta w$ as well as $J_a$. As a result, 
Eq.~\eqref{eq:Jdot} does not vanish. Furthermore, 
$\left\langle H^{(1)}\right\rangle$ depends on the initial phase only through the combination $\Delta w$, which implies 
\begin{align}
 \left\langle \frac{dJ_a}{d\tau}\right\rangle
 =-\frac{\partial \left\langle H^{(1)}\right\rangle}{\partial \Delta w}n_a\,,  
\end{align}
where we have defined $n_a\equiv \{m\Omega,n,k,m\}$. The above relation immediately explains why the jumps of ``constants of motion'' vanish when $k+m=$ odd
as follows (see also the previous discussions~\cite{Gupta:2022fbe,Silva2022}).
Under the parity transformation $\vec x\to -\vec x$, the tidal field remains invariant. 
However, the angle variables transform like $\{w^t, w^r, w^\theta, w^\phi\} \to \{w^t, w^r, w^\theta+\pi, w^\phi+\pi \}$. 
Therefore, under the parity transformation, $\Delta w$ transforms as $\Delta w\to \Delta w + (k+m)\pi$ indicating that  $\left\langle H^{(1)}(\Delta w)\right\rangle=\left\langle H^{(1)}(\Delta w+\pi)\right\rangle$ when $k+m$ is odd. 
Further, $H^{(1)}$ can be expanded as a Fourier series of 
$w^a$ as
\begin{align}
 H^{(1)} =\sum_{n,k,m} e^{i n_a w^a} H^{(1)}_{n,k,m}\,. 
 \label{eq:Hexpansion}
\end{align}
From the expression~\eqref{eq:Hexpansion}, we find that the long-time average of $H^{(1)}$, 
\begin{align}
 \left\langle H^{(1)}\right\rangle =\sum_{s\in \mathbb{Z}} e^{i s \Delta w} H^{(1)}_{sn,sk,sm}
\end{align}
is a periodic function with respect to $\Delta w$ with a period of $2\pi$. 
Let us focus on the dominant contribution coming from $s=\pm 1$.  
Then, if $\left\langle H^{(1)}\right\rangle$ takes a maximum value at $\Delta w$, 
it should take a minimum value at $\Delta w+\pi$.  
Therefore, we can conclude that the function $\left\langle H^{(1)}(\Delta w)\right\rangle$ restricted to the contribution from 
$s=\pm 1$ becomes 
constant for $k+m=$ odd. Nonetheless, it is possible to have the contribution coming from the terms with $s=$ even in the resonance with $k+m=$ odd. Such higher-order contributions, which have not been specifically discussed in literature~\cite{Gupta:2022fbe, PaperI}, do not necessarily vanish. To provide more complete fitting formulae for resonance jumps, we will return to this point in our future publication.

So far, we did not consider the lowest order (adiabatic) self-force, {\it i.e.}, the secular change of the ``constants of motion'' caused by the linear perturbation induced by the particle itself. In the actual evolution of EMRIs, this adiabatic self-force is expected to dominate the evolution of the ``constants of motion'', while the tidal resonance gives just minor corrections. 
In such cases, we can compute the jumps, {\it i.e.}, 
the extra changes induced in the ``constants of motion'' due to a tidal resonance by integrating along the evolution caused by the adiabatic self-force across the resonance point. At least, in this approximation we obtain the relations

\begin{eqnarray}
\label{eq:resJumpRel}
\centering
\hspace{-2cm}
\Delta J_t(=-\Delta E):\Delta J_r:\Delta J_\theta: \Delta J_\phi(=\Delta L_z) &=&
\left\langle \frac{d J_t}{d\tau} \right\rangle : \left\langle \frac{d J_r}{d\tau} \right\rangle : \left\langle \frac{d J_\theta}{d\tau} \right\rangle : \left\langle \frac{d J_\phi}{d\tau} \right\rangle 
\nonumber \\
&=& 
m\Omega:n:k:m.
\end{eqnarray}
This is the main message of this short paper. To understand the resonant jump, it is sufficient to compute only one of $\Delta J_a$, {\it e.g.}, $\Delta L_z$. 
An exception are $m=0$ resonances, for which $\Delta L_z=0$ is obvious due to axisymmetry of $m=0$ perturbation and therefore does not convey information about other jumps.


\section{Demonstration of the validity of the relation among resonance jumps}
In the following, we demonstrate the validity of the relation shown in Eq.~\eqref{eq:resJumpRel}. It appears that this equation gives three independent relations among $\Delta J_a$. One is just the conservation law coming from the presence of Killing vector, {\it i.e.}, 
$dJ_t/d\tau-\Omega dJ_\phi/d\tau=0$, which reduces to the familiar relation $dE/d\tau=0$ in 
the static limit. 
Another is the conservation of $\mu^2\equiv -2H^{(0)}$, meaning that there is only one independent relation.
The full Hamiltonian $H$ is conserved under the influence of any kind of perturbation to the background space-time. 
Hence, $H^{(0)}$ can vary in time but only within the magnitude of the variation of $H^{(1)}$. 
Therefore, $H^{(0)}=-2\mu^2$ is not allowed to violate indefinitely in a secular manner. After a long time average for any kind of perturbation, we should have 
\begin{align}\
  0=\left\langle \frac{d H^{(0)}}{d\tau}\right\rangle
  =\left\langle \frac{\partial H^{(0)}}{\partial J_a}\frac{d J_a}{d\tau}\right\rangle
     =\omega^a \left\langle \frac{d J_a}{d\tau}\right\rangle\,. 
\end{align}
If we substitute the relation~\eqref{eq:resJumpRel}, we find that the vanishing 
of the right-hand side is nothing but the resonance condition itself. 
Hence, only one of three independent relations derived from Eq.~\eqref{eq:resJumpRel} 
is the non-trivial one that must be satisfied by generic perturbation. 

Next, to obtain the formula for the ratio $\langle dQ/d\tau\rangle/\langle dL_z/d\tau\rangle$, 
it is easier to consider that $Q$ is just the function of $\{E,L_z,J_r\}$ by setting 
$\mu^2=1$. In this treatment, $J_\theta$ is a variable depending on $\{E,L_z,J_r\}$. 
Then, we have 
\begin{align}
 1=\left(\frac{\partial Q}{\partial Q}\right)_{\!\!E,L_z}=
   \left(\frac{\partial Q}{\partial J_r}\right)_{\!\!E,L_z}
   \left(\frac{\partial J_r}{\partial Q}\right)_{\!\!E,L_z}\,,
\end{align}
and 
\begin{align}
 0=\left(\frac{\partial Q}{\partial L_z}\right)_{\!\!E,Q}=
   \left(\frac{\partial Q}{\partial J_r}\right)_{\!\!E,L_z}
   \left(\frac{\partial J_r}{\partial L_z}\right)_{\!\!E,Q}
   +  \left(\frac{\partial Q}{\partial L_z}\right)_{\!\!E,J_r}\,. 
\end{align}
From these relations, we immediately obtain
\begin{align}
  \frac{dQ}{d\tau} &= \left(\frac{\partial Q}{\partial J_r}\right)_{\!\!E,L_z} \frac{dJ_r}{d\tau}
     + \left(\frac{\partial Q}{\partial L_z}\right)_{\!\!E,J_r} \frac{dL_z }{d\tau}\cr
    & =\left[\frac{n}{m}-\left(\frac{\partial J_r}{\partial L_z}\right)_{\!\!E,Q}\right] 
       \left(\frac{\partial J_r}{\partial Q}\right)_{\!\!E,L_z}^{-1}\frac{dL_z}{d\tau}\,,
       \label{eq:JrExpression}
\end{align}
where in the second equality we used the relation \eqref{eq:resJumpRel}. 
In the same manner we can obtain an equivalent expression in terms of $J_\theta$, which is given by 
\begin{align}
 \frac{dQ}{d\tau} =\left[\frac{k}{m}-\left(\frac{\partial J_\theta}{\partial L_z}\right)_{\!\! E,Q}\right] 
       \left(\frac{\partial J_\theta}{\partial Q}\right)_{\!\!E,L_z}^{-1}\frac{dL_z}{d\tau}\,.  
   \label{eq:JthExpressionpre}
\end{align}
The above relation can be more explicitly written as
\begin{align}
 \frac{dQ}{d\tau} =\frac{\displaystyle\frac{\pi a k}{m}\sqrt{1- E^2}\, y_+ -{2 L_z}\left(K\left[\left(\frac{y_-}{y_+}\right)^2\right]-\Pi\left[y_-^2,\left(\frac{y_-}{y_+}\right)^2\right]\right)} 
       {\displaystyle{K\left[\left(\frac{y_-}{y_+}\right)^2\right]}}\frac{dL_z}{d\tau}\,.  
   \label{eq:JthExpression}
\end{align}
where $y_{\pm}~ (y_+>y_-)$ are the values of $\cos^2\!\theta$ at zeroes of the potential $V_\theta$ and $K[m], \, \Pi[n,m]$ are complete elliptic integrals of the first and third kind, respectively.
Explicit expressions for all the quantities are given in~\ref{appex:A}. 

\begin{figure*}
\begin{center}
  \includegraphics[width=0.4\linewidth]{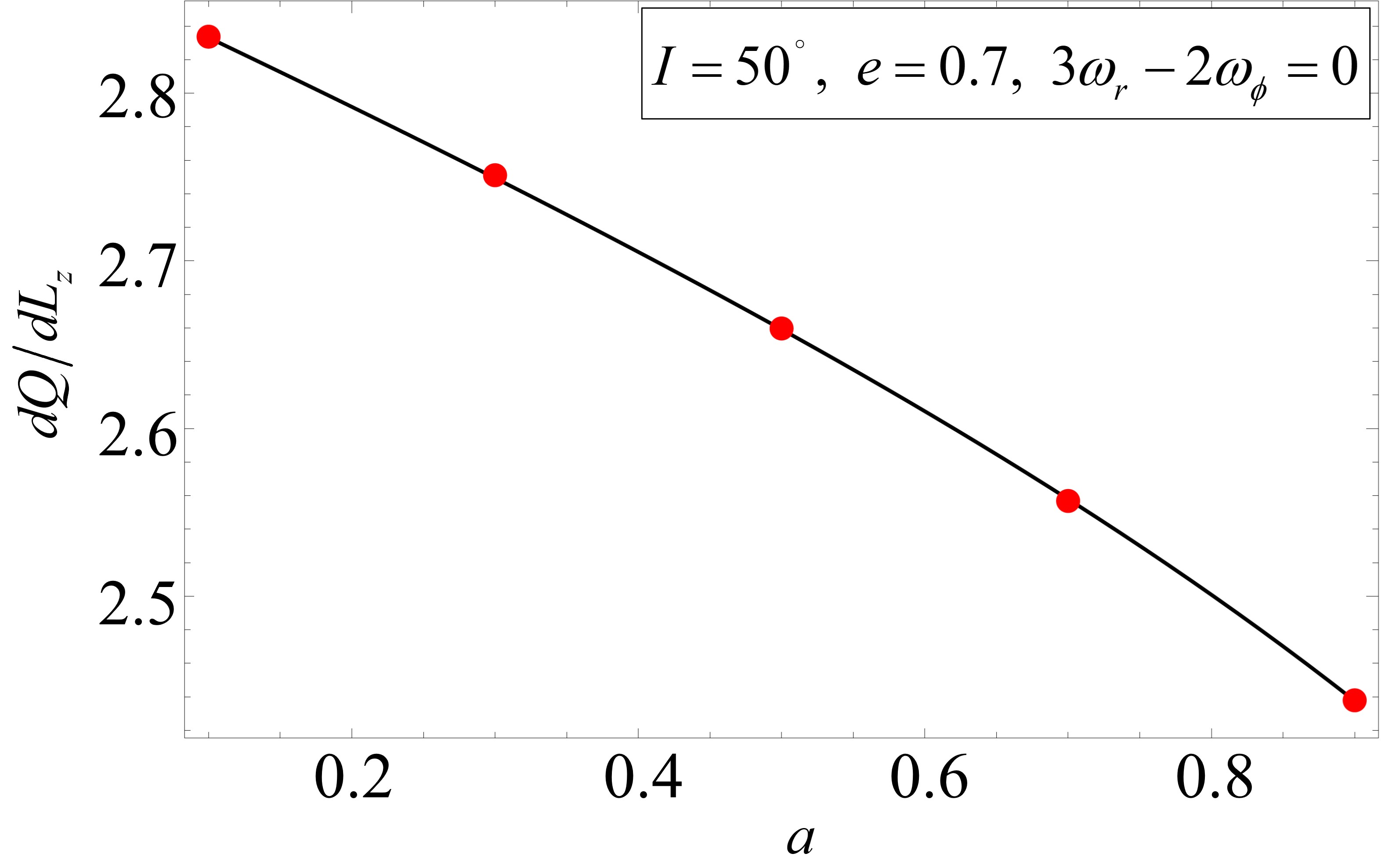}
    \hskip 1cm
\includegraphics[width=0.4\linewidth]{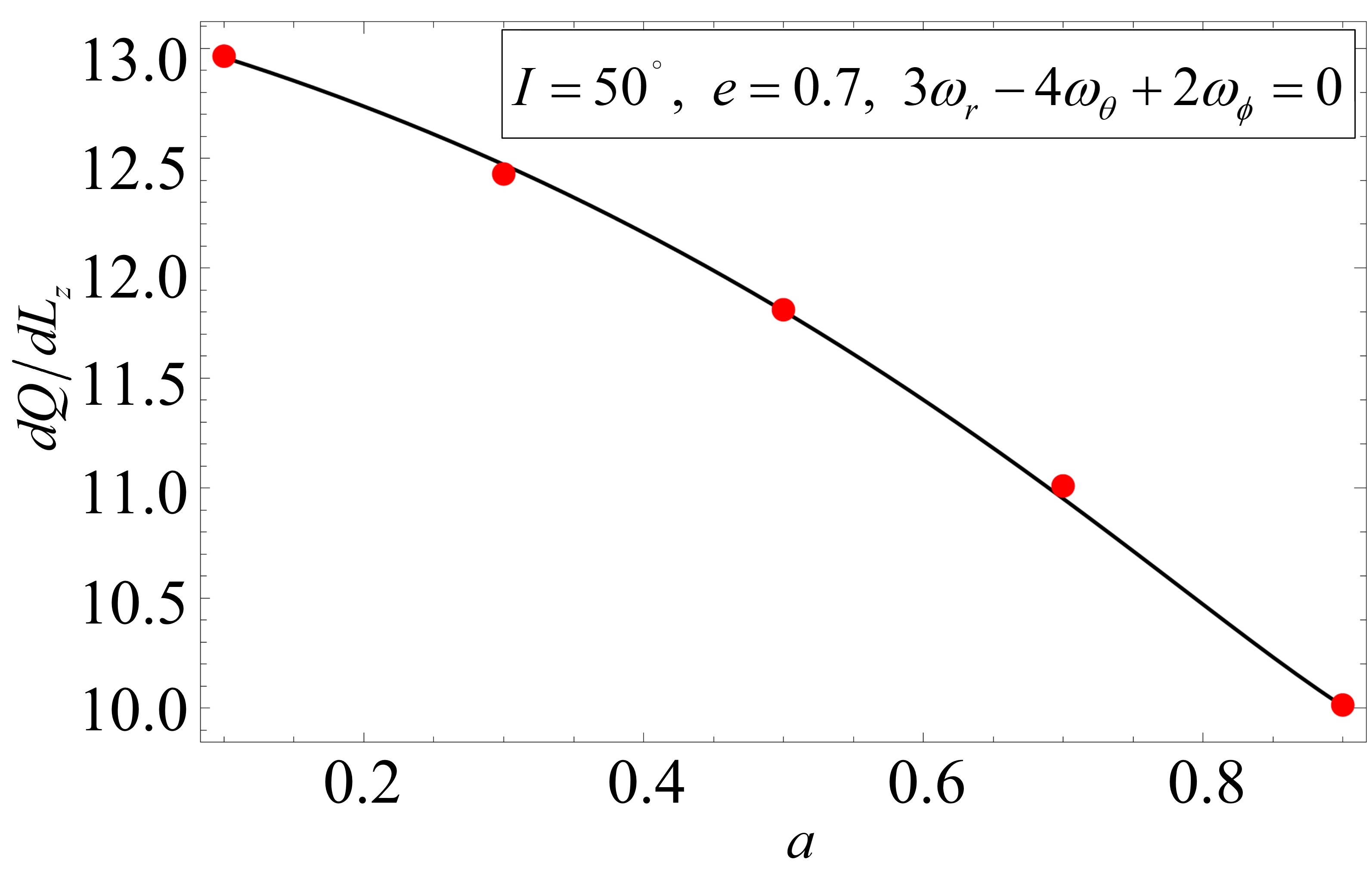}
\hskip 0.5cm
\vskip 0.6cm
\includegraphics[width=0.4\linewidth]{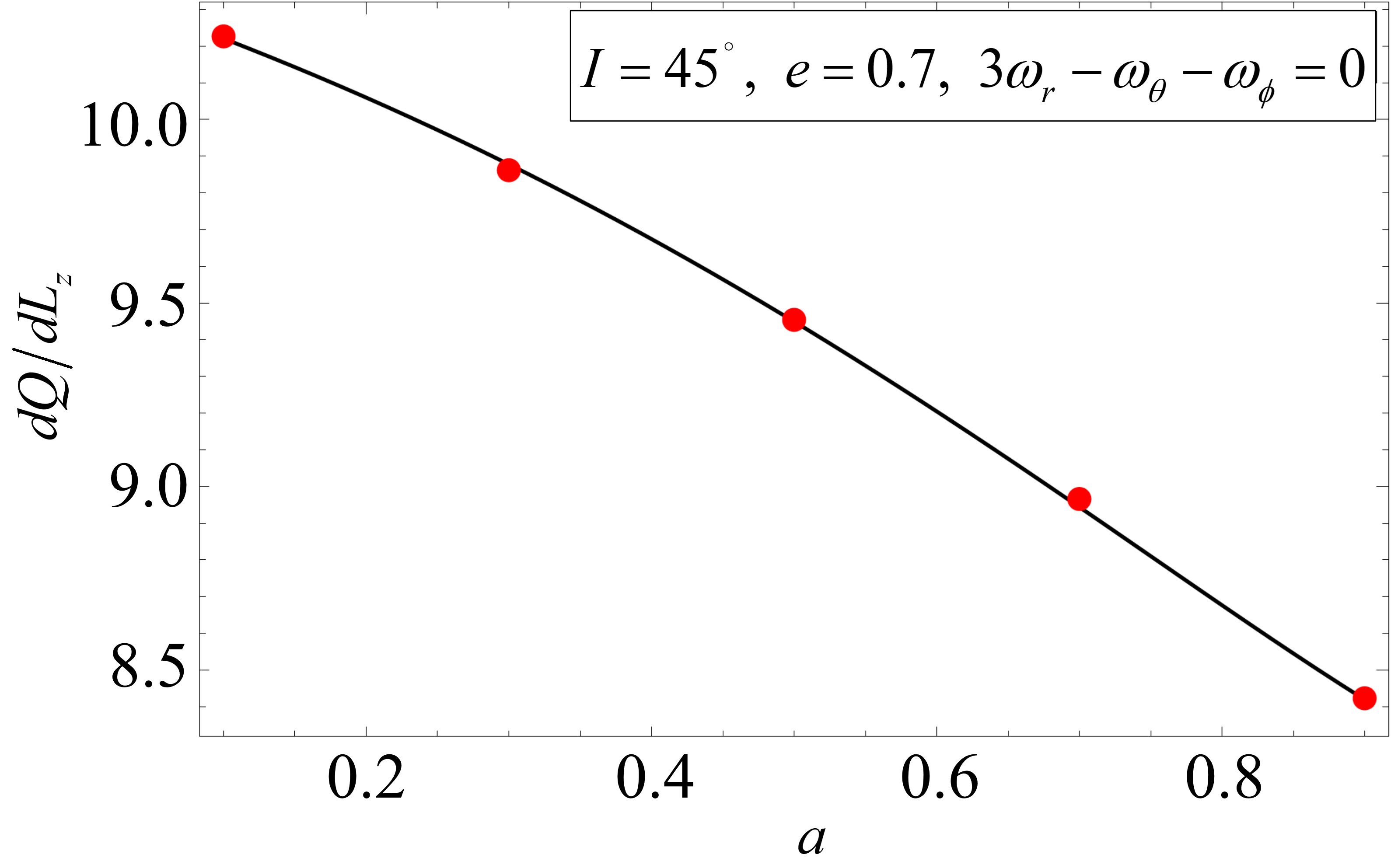}

\caption{Dependence of the ratio of the average change rate of Carter constant and axial angular momentum on the spin of the central BH. The figures are plotted for three different resonance combinations: {$3:0:-2$} (top-left), {$3:-4:2$} (top-right), {$-3:1:1$} (bottom). The perturber is placed on the equatorial plane in the case of $m =\pm 2$ resonance, whereas in the case of $m =\pm 1$ resonance, it is positioned at an inclination of $45^\circ$. The dots represent the values obtained from the relation given in Eq.~\eqref{eq:JthExpression} and curves denotes the fitting given in Ref.~\cite{PaperI,Gupta:2022fbe}}.
\label{fig:resonance_compare}
\end{center}
\end{figure*}

Now, we confirm Eq.~\eqref{eq:JthExpression} using the fitting formulae derived in Ref.~\cite{Gupta:2022fbe}. 
These formulae are derived under the assumption of a static perturber with $\Omega=0$. 
Figure \ref{fig:resonance_compare} shows an agreement for different resonance combinations such as $n:k:m=3:0:-2$, $n:k:m=3:-4:2$, and $n : k : m = -3 : 1 : 1$. The horizontal axis is the spin of the central black hole and orbital eccentricity ($e$) and inclination ($I$) is fixed (see the figure caption for details).
We find that the fitting formulae satisfy the relation \eqref{eq:JrExpression} or \eqref{eq:JthExpression} within the error of 0.01\%. 

\section{Discussion}
Our work establishes a simple relation between the jump of the Carter constant and that in the angular momentum caused by the tidal resonance effect, derived from the fact that the interaction Hamiltonian due to the tidal perturbation depends on the time independent combination of the orbital phases at the resonance. 
The ratio between 
the change rate of the Carter constant and that of 
the the angular momentum is algebraically given by the background values of the ``constants of motion". 
Hence, the jump of the Carter constant across the resonance point
can be easily calculated from the jump of the angular momentum. 

Additionally, we validated this new relation by comparing them with the fitting formula provided in Ref.~\cite{PaperI,Gupta:2022fbe}, which is also a consistency check of our previously obtained fitting formulae. However, even if there were some mistake in the expressions for the metric perturbation adopted to compute the tidal force, this proportionality relation is still valid as long as $dJ_a/d\tau$ is correctly calculated from the same metric perturbation. This means that the applicability of this proportionality relation is not limited to the case of tidal resonance. 
An interesting application would be the evaluation of the resonance jumps in 
the context of non-GR modification to the background Kerr geometry~\cite{Lukes-Gerakopoulos:2010ipp,Gair:2007kr}.
We hope to return to this issue in our future publication. 

In contrast to the tidal resonance case, we cannot use this relation in evaluating the conservative part of the self-force resonance, because the self-force resonance appears only when the resonance condition with $m=0$ is satisfied, and hence the jump of the angular momentum trivially vanishes. 
The current discussion does not apply to the dissipative part of the self-force, either. In this case, to compute $\dot J_a$, the variation of the interaction Hamiltonian $H^{(1)}$ with respect to the phase variables $q^a$ should be taken first before setting the initial phases of the source orbit to those of the argument of the perturbation field. Afterward, the long-time average of the expression for $\dot J_a$ is taken.  
Therefore, $\dot J_a$ is not given by a simple variation of a function of the initial phases. 

\ack
 We thank Soichiro Isoyama for valuable comments as well as B\'eatrice Bonga for feedback on our draft. This work was supported by Japan Society for the Promotion of Science (JSPS) Grants-in-Aid for Scientific Research (KAKENHI) Grant Numbers JP17H06357, JP17H06358, JP20K03928 (T.T.). PG is supported by JSPS fellowship and KAKENHI Grant Number 21J15826.

\appendix
\section{Explicit expression}
\label{appex:A}

Here, we give closed form expressions for ${\partial J_\theta}/{\partial L_z}$ and ${\partial J_\theta}/{\partial Q}$ requisite for Eq. \eqref{eq:JthExpression}.

\begin{align}
\frac{\partial J_\theta}{\partial L_z}
&=
\frac{2 L_z}{\pi a \sqrt{1-E^2}}\int_0^{y_-}\frac{dy}{\sqrt{(y_-^2-y^2)(y_+^2-y^2)}}
\left\{1-\frac1{1-y^2}\right\}\cr
&=
\frac{2 L_z}{\pi a \sqrt{1-E^2}\, y_+}\left(K\left[\left(\frac{y_-}{y_+}\right)^{\!2}\right]-\Pi\left[y_-^2,\left(\frac{y_-}{y_+}\right)^{\!2}\right]\right)\,,
\label{eq:A1}\\
\frac{\partial J_\theta}{\partial Q}
&=
\frac{1}{\pi a \sqrt{1-E^2}}\int_0^{y_-}\frac{dy}{\sqrt{(y_-^2-y^2)(y_+^2-y^2)}}\cr
&=\frac{1}{\pi a\sqrt{1-E^2}\, y_+}K\left[\left(\frac{y_-}{y_+}\right)^{\!2}\right]\,,
\label{eq:A2}
\end{align}
where $K[m]$ and $\Pi[n,m]$ are complete elliptic integral of the first and of the third kind,  respectively, defined by
\begin{align}
&K[m]\equiv\int_0^{\frac{\pi}{2}} \frac{d\phi}{\sqrt{1-m \sin^2\! \phi}}
=\int_0^1 \frac{dx}{\sqrt{(1-x^2)(1-m x^2)}}\,, \\
&\Pi [n,m]\equiv\int_0^\frac{\pi}{2} \frac{d\phi}{(1-n \sin^2\!\phi)\sqrt{1-m \sin^2\!\phi}}
=\int_0^1 \frac{dx}{(1-n x^2)\sqrt{(1-x^2)(1-m x^2)}}\,, 
\end{align} 
and $y^2_\pm$ are given by 
\begin{align}
 y_\pm^2 = \frac {a^2 \left (1 - E^2 \right) + L_z^2 + 
   Q \pm \sqrt {\left (a^2\left (1 - E^2 \right) + L_z^2 + 
          Q \right)^2 + 
      4 a^2\left (E^2 - 1 \right) Q}} {2 a^2\left (1 - 
     E^2 \right)}\,,
\end{align}
which are the values of $\cos^2\!\!\theta$ at the zeros 
of the potential $V_\theta$. Since $1-E^2>0$ and $Q>0$, we find $y_+>y_->0$.
In deriving Eqs.~\eqref{eq:A1} and \eqref{eq:A2}, we have also used the fact that the potential is written as 
\begin{align}
V_\theta(\theta)=a^2 (1-E^2) \frac{(y_+^2-\cos^2\!\theta)(y_-^2-\cos^2\!\theta)}{\sin^2\!\theta}\,. 
\end{align}

\section*{References}
 
\bibliographystyle{unsrt} 

\end{document}